# Copper Doped Zinc Oxide Nanocrystalline Thin Film: Growth, Characterization, Effect of annealing on Photocatalytic activity and electrical properties of irradiated film

Mohana F. Attia

mohana.attia@alasala.edu.sa

Physics, Engineering College, Alasala National Colleges, Saudi Arabia.

**Abstract** The problem of detecting the ability of some metal oxides as photocatalysts is receiving considerable attention in the world. The studied ZnO is considered as one of the most utilized semiconductors, due to its direct band gap of 3.37 eV and a large exciton binding energy of 60 meV at room temperature. The characteristic formulation of Cu - doped ZnO is implemented to obtain an improved adsorption kinetics and photocatalytic activity of ZnO photocatalyst capable of evolving estimated photoluminescence (PL) spectra of ZnO nanoparticles to explore the effect of Cu-doping and annealing temperature on its optical properties. using Sol – gel spin coating technique, Zinc oxide and copper doped zinc oxide thin films were deposited on borosilicate substrate and all films was annealed at 500 C for 5- hours. Optical properties of the films have been studied using spectrophotometer in the wavelength range 300-800nm and structure of ZnO films is investigated and included reported, also the effect of the doping concentration of Copper. Further the characteristic formulation to treat the photocatalytic performance of pure ZnO and Cu - Doped ZnO samples were investigated through the degradation of Malachite Green (MG), as a model of organic pollutants, under UV light irradiation. Mainly, these films exhibited a good transparency in the visible region. It was observed that when the doping concentration increased, the transmittance also increased. All the prepared catalysts mainly showed two emission regions: a sharp peak in the ultraviolet region and another broad peak in the visible region. The photocatalytic activity was achieved by the degradation of malachite green (MG) aqueous solution under UV irradiation. The findings showed that the increased annealing of Cu doped ZnO with CuO on the surface resulted in highly improved photocatalytic activity. In summary, catalysts activity was influenced by both Cu different ratio loading on ZnO and annealing temperature. The effect of nitrogen and electron beam irradiation on the electrical properties of films. Films of 3 mm were exposed to a charged (ion/electron) beam for different treatment times (20, 40, and 80 minutes); the beam was produced from a dual beam source using nitrogen gas with the other ion/electron source parameters optimized. The real ($\varepsilon'$) and imaginary ($\varepsilon''$) parts of the dielectric constant decreased with frequency for all irradiated and non-irradiated samples. The AC conductivity showed an increase with frequency for all samples under the influence of both ion and electron irradiation for different times.

## I. INTRODUCTION

Zinc oxide (ZnO) is a highly good material for many varieties of applications due to its minimal fabrication cost, environmentally friendly, compatibility and simple fabrication process. ZnO crystallizes in a hexagonal wurtzite structure [1]. with wide direct band gap [2] and potent exciting binding energy of 60 meV [3]. doping ZnO with a transition metal such as Cu [4] has been verified as an effective method to adjust its functionality including electrical and optical properties Cu-doped ZnO has shown significantly improved properties such as electrical, magnetic, photocatalytic performance and gas sensing properties. In practice, ZnO-based thin films can be grown by diverse growth techniques including radio frequency magnetron sputtering (RFMS), pulsed laser deposition (PLD), laser molecular beam epitaxy (P-MBE) [5], spray pyrolysis [6], metal organic chemical vapour deposition (MOCVD) [7], and sol-gel spin coating method [8]. Among these methods, spin coating can provide the ease of chemical composition of doping, which is an advantage over the others. Radiation handling of polymer materials lid undergoing polymers to irradiation, generally in a continuous mode, to modify the polymers to better their characteristics for industrial purposes. Radiation handling of polymers is a non-power application and at most consists of cross linking, curing, grafting, and degradation. Ion beam irradiation is a well decided tool for the modification of polymer surfaces, useful for **controlled changes of a variety of characteristics, like chemical reactivity, hardness, wear** [9-11] and electron beam effects on polymers [12] have obtained significance in recent years, in view of the probable applications, such as in surface science. Electron beam irradiation is a based agent for modifying the chemical structure and physical characteristics of a polymer's surface, such as adhesion, friction, wetting, swelling, dielectric characteristics, and biological compatibility [13]. In this study, the structural and optical properties of the Cu-doped ZnO(CZO) thin films prepared by sol-gel spin coating were investigated. Films of 3 mm were exposed to a charged (ion/electron) beam for different treatment times (20, 40, and 80 minutes). The dielectric loss tangent tan δ, electrical conductivity σ, and dielectric constant $\varepsilon^{'}$ in the frequency range 100 Hz–100 kHz were measured at room temperature. The variation of dielectric constant and loss tangent as a function of frequency was also studied at room temperature.

## II. MATERIAL AND METHODS

### A. Research duration, period and location of study

The research was carried out during the September 2017 to march 2018. It comprised the 6 month at Alasala National Colleges, Saudi Arabia.

### B. Materials

Zinc acetate dihydrate [Zn (CH3 COO)$_2$ .2H$_2$O], absolute methanol 2.3.dihydroxysuccinic acid (CHOH-COOH)$_2$, Cupper nitrate trihydrate [Cu(NO$_3$)$_2$ .3H$_2$O] and malachite green(MG). All chemicals were utilized without further purification. Double distilled water was used in all solution preparations.

### C. Methods

Two Synthesis of ZnO were designed to study the growth, characterization of copper doped zinc Oxide and effects of annealing on photocatalytic activity.

1. Synthesis of pure ZnO and Cu- doped ZnO nanoparticles

Cu – doped ZnO thin films were prepared by sol-gel method using zinc acetate dehydrate (CH$_3$ COO)$_2$ Zn·2H$_2$O and Copper acetate dihydrate Cu$_3$ [(CH$_3$ COO)2·H$_2$O] as starting precursors. The 2-methoxy ethanol (C$_3$H$_8$O$_2$) and diethanal amine (DEA) were selected as the solvent (0.5 M, 100 mL) and sol stabilizer, respectively. The prepared mixture was vigorously stirred at 100 °C for 5 h by magnetic stirrer and cooled to room temperature for 24 h. The precursors prepared at different copper concentration of 2-20 wt.% were spin-coated on borosilicate substrates at room temperature with speed of 3000 rpm for 30 s. After repeating the coating procedure 6-times, all coated films were annealed at various temperatures ranging from $550^0$ C for 3 h in ambient air.

2. Synthesis of pure ZnO and Cu-doped ZnO nanoparticles

A certain amount of zinc acetate dihydrate was dissolved in 75 mL of methanol, after stirring for 50 min, an appropriate amount of copper nitrate trihydrate with a certain molar ratio Cu/Zn was also added and the mixture was stirred for 50 min. Further, an amount of double distilled water solution (50 mL) of tartaric acid was added dropwise to the previous mixture. The gel formed rapidly [10 minutes] at room temperature under vigorous magnetic stirring and dried at 100°C for 10 h. Subsequently, the dry gel was first grinded and annealed at 550°C for 1 h. The pure ZnO nanomaterial was synthesized following the same procedure without adding copper nitrate trihydrate.

### D. Irradiation

Different concentrations (2, 4, and 6 g) were dissolved to obtain uniform and homogenous surfaces. It was found that, the best films are 0.03 and 0.05 g/ml ones. After preparing the film, it was exposed to different treatment times (10, 25, and 40 minutes) from a dual ion/electron beam source with low energy using nitrogen gas. The electrical properties of the polymer samples were investigated at room temperature by using a system [14]. DSP Dual Phase Lock-In Amplifier (SR830) is used to measure the voltage difference (VR) between the two ends of known resistance R in an equivalent RC circuit. Measurements were carried out at frequency range (1mHz–102.4 kHz) over the temperature range (200–400 K) in vacuum of about $1 \times 10^{-5}$ mbar. Data logger software is used for controlling and gathering the data through IEEE-488 GPIB and 332 serial interfaces with different tools. The two surfaces of each polymer sample were coated with silver paint, checked for good conduction, and then kept in between the two cell electrodes for making measurements. The 3 mm thin films were exposed to a charged (ion/electron) beam for various treating times (0, 20, 40, and 80 minutes); the radiation was generated from a dual radiation source using nitrogen gas with the other ion/electron source factors optimized.

The experimental framework used for dual beam treating is a local built system, shown in Figs. 1(a) and (b) [15]. The source consists of two copper anode rods surrounded by a cylindrical cathode. A blank Perspex cylinder was put inside the cylindrical cathode, relative to it. The Perspex cylinder had two opposite holes with a diameter of 5 mm, to output the highest ion and or electron beam current. The film samples were put on a Faraday cup at both sides 1.5 cm from the cylindrical cathode. A saddle-field ion source was used to obtain the dual beam with nitrogen and consequently irradiate the film surface. In case of the ion source, the anode was connected to the positive polarity of the power supply, while, in case of the electron source, the anode was connected to the negative polarity and the cathode was earthed. The system was evacuated up to about $5 \times 10^{-5}$ mbar to remove the residual gases before gas injection into the source.

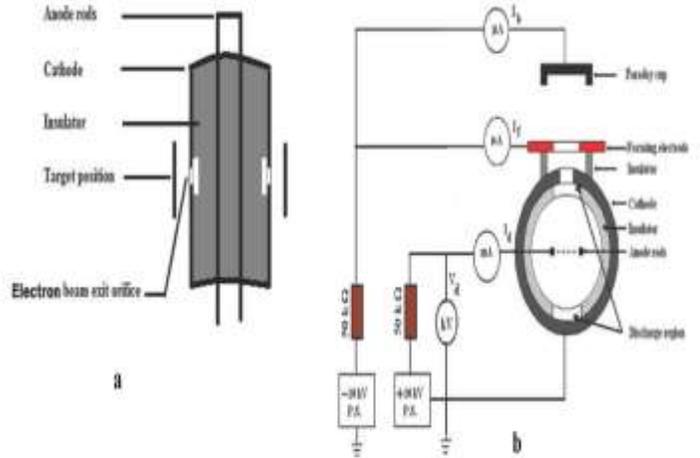

Figure 1: Schematic of the double beam source, a. Electron source and b Ion source [16]

The ion/electron fluence was estimated by time of irradiation and beam current as [17]:

$$I = Q/t = Dqe/t = \varphi\, Aqe/t \qquad (1)$$

Where $I$ is the ion current (A), $Q$ the total charge, D the dose (ion fluence in ion/cm² × area) of irradiation in $cm^2$, $q$ the charge state, e the electron charge (1.6 × 10⁻¹⁹ C), and $t$ the irradiation time in $s$.

The conductivity can be expressed according to [18]:

$$\sigma(\omega) = \sigma_{dc}(0) + \sigma_{ac}(\omega) \qquad (2)$$

Where $\sigma_{dc}(0)$ is the DC conductivity, $\sigma_{ac}(\omega)$ is the AC conductivity, and $\omega$ is the frequency.

The dielectric constant ($\acute{\varepsilon}$) measurements were carried out in a frequency range from 100 Hz to 100 kHz at room temperature.

The dielectric constant was determined from the formula:

$$\sigma = I \cos\theta\, (d/A),\; \acute{\varepsilon} = \frac{cd}{\varepsilon_0 A},\; \varepsilon'' = \tan\delta\, \acute{\varepsilon} \text{ and}$$

$$\tan\delta = \frac{1}{\tan\theta} \qquad (3)$$

where *c* is the capacitance in Farad, *d* is the thickness of the sample in m, *A* is the cross-sectional area, $\varepsilon_0$ is the constant of permittivity of free space, and *θ* is the phase angle. Electrical measurements were carried out on the film samples in the form of a disc with polished surfaces covered by silver paste and with dimensions of 1 cm in diameter and 3 mm in thickness:

$$\tan\theta = \frac{1}{2\pi f R_p C_p} \quad (4)$$

Where *δ* is the loss angle, *f* is the frequency, **Rp** is the equivalent parallel resistance, and **Cp** is the equivalent parallel capacitance.

The dielectric loss $\varepsilon''$ was also measured in terms of the tangent loss factor (*tan δ*), defined by the relation:

$$\varepsilon'' = \varepsilon' \tan\delta \quad (5)$$

### E. Materials characterization

The morphology of the nanopowder samples was examined using a scanning electron microscope. The samples were previously oven dried and coated with a thin film of gold to provide ZnO powder surface with electrical conduction. The composition and average size of nanoparticles were determined by the powder X-ray diffraction patterns, the samples were recorded by a diffractometer Bruker D8 - Advance, measurements were performed to identify the structural properties and crystalline behavior of the films using Cu-Kα radiation (λ =0.15406 nm). The accelerating voltage and scanning angle were 40 kV and $20^0$ –$75^0$c, respectively. The chemical bonding and formation of wurtzite structure in ZnO (pure) and Cu-doped ZnO were confirmed by Fourier transform infrared spectroscopy (FTIR) measurements at room temperature. The surface morphologies were examined by scanning electron microscope (SEM). The optical measurements of the Cu – doped ZnO thin films were carried out at room temperature using spectrophotometer [Labomed – UVS 2800] in the wavelength range of 190 nm to 1100 nm. Room temperature photoluminescence studies were carried out using the (Perkin Elmer LS55 Luminescence).

### F. Determination of heterogeneous photocatalytic activity

The photoactivity of purely prepared ZnO and Cu-doped ZnO were examined using (MG) dye as a pollutant and all experiments were carried out in a Pyrex photoreactor under U.V irradiation. After achieving adsorption equilibrium in the dark, the solution was illuminated for photocatalytic kinetic study: The samples of the MG solution were taken after different irradiation times and were analyzed using UV–visible spectroscopy at a wavelength max=618 nm. Using the Beer-Lambert law, the absorption measurement was converted to concentration. The photocatalytic degradation efficiency was calculated using the following equation:

Mg % = $[(C_0 – C_t)/C_0] \times 100\%$ (6)

Where $C_0$ is the [MG] initial concentration, and $C_t$ is the [MG] concentration at time *t*.

## III. RESULTS AND DISCUSSION

### 1. Scanning electron microscopy image (SEM)

The SEM micrographs of pure ZnO and Cu - ZnO annealed at 550 $^0$C was shown in Fig.1. The particles of the obtained powder were spherical in shape.

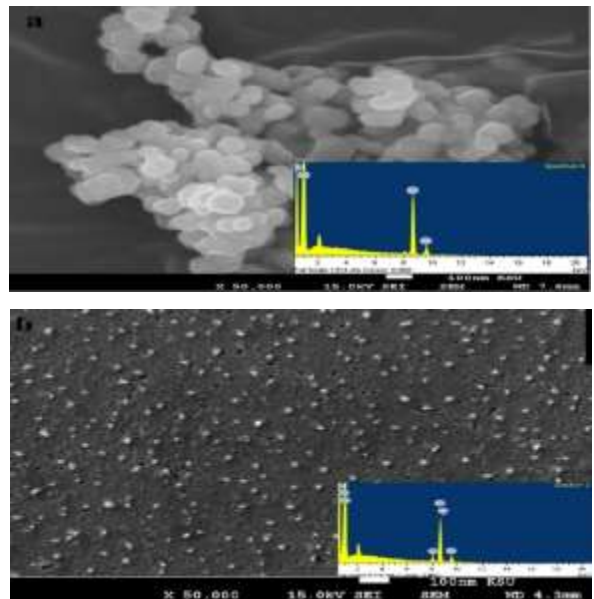

Fig.1: SEM micrographs of ZnO films annealed at (a) $550^0$c, (b) 20 wt. % Cu doped ZnO film.

It was clearly observed that both films have a smooth surface comprising uniform grain size. As the annealing temperature increases, the crystalline structure and grain size of the film improves. The morphology of the 20 wt. % CZO show mixtures of Nanocrystalline CZO grains and a big number of other granular particles which may be CuO particle due to redundant Cu doping. This characteristic suggested the limitation of Cu doped on ZnO film. The random tendency of grains and the uneven surface may be related to the growth being along different crystal orientation.

2. *Characterization of the prepared catalysts by X-ray diffraction (XRD)*

In the X-ray diffraction patterns of ZnO and 20 wt.% Cu doped ZnO films with various coating repetitions. The patterns were shown in Fig. 2 a, b corresponds to three main diffraction peaks of crystallized ZnO, namely (100), (002) and (101) positioned at $2\theta = 31.6°$, $34.3°$, and $36.1°$ respectively. This result suggested that the as-prepared films annealed at 550 C have polycrystalline films with a hexagonal wurtzite structure. In addition, the appearance of a very low intensity diffraction peak at position 38.60 corresponding to CuO as the monoclinic (base-centered) phase (JCPDS#18-1916) was observed. The films did not show any preferential orientation of crystallization accompanying almost same intensities of the peaks (100), (002) and (101). The peak intensities were very weak because the thickness of the film is quite thin. Although the peak intensities are weak, it can be easily shown that the crystallinity of the ZnO film deteriorates with Cu incorporation [19]. The average crystallite size of all films can be calculated from the full width at half maximum (FWHM) and angular position of the (002) diffraction peak by Scherrer's formula:

$$D = \frac{(K\lambda)}{(\beta \cos\theta)} \quad (7)$$

Where, $D$ is the grain diameter, $K$ is the shape factor, $\lambda$ is the X-ray wavelength of Cu K$\alpha$ (0.154 nm), $\beta$ is the full-width at half maximum (FWHM) and $\theta$ is the Bragg angle [20].

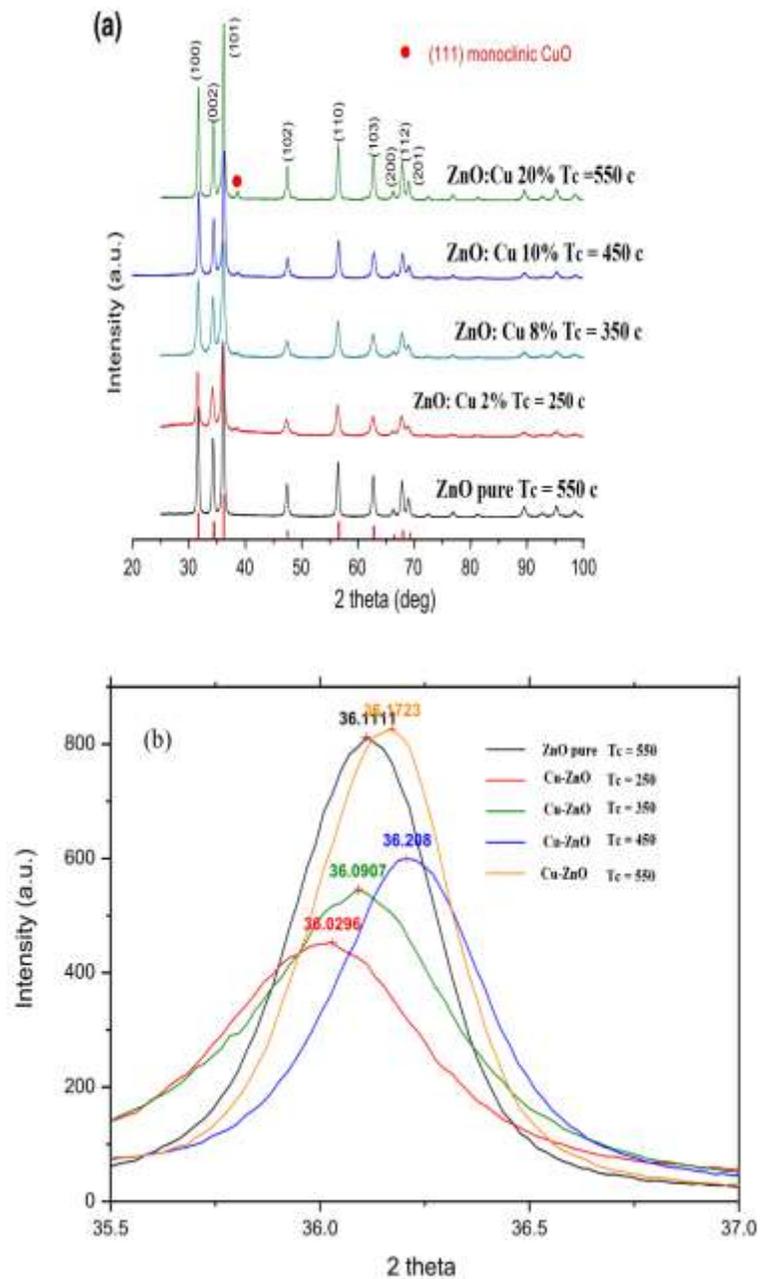

Fig.2: **a**: X-ray diffraction patterns of ZnO and Cu-doped ZnO films with different thickness and Tc.
**b**: XRD principal peak shift due to the effects of the Cu doping and Tc.

The effect of Cu doping can be observed at the same annealing temperature ($T_C = 550^0$ C) as shown in Fig. 2. In comparison with pure ZnO, the introduction of copper ions shifted the diffraction peaks to higher angles by 0.060. Indeed, this change can be attributed to the replacement of Zn ions (ionic radii = 0.60 A) by Cu ions (ionic radii = 0.57 A) or may be due to structural stresses and modification of the lattice parameters of ZnO. This shift value indicated a decrease in the lattice parameters since the value of c parameter decreased by about 0.01 A from undoped ZnO (5.226 A) to Cu-doped ZnO (5.217 A). This decrease confirmed that the substitution of Zn ions by Cu in the ZnO lattice, was facilitated by the similarity of the ionic radii. Sharma et al. [21], in their study on Mn-doped ZnO, reported similar observations.

The annealing temperature ($T_C$) effect was observed on both intensities of XRD peaks and lattice parameters as shown in Fig. 2. The principal peak (101) intensity increased with increase in $T_C$, and the shape of this peak became narrower, indicating an increase in particle size. On the other hand, the effect of $T_C$ on the lattice parameters showed the shift of the peak at 2h = 36.11, which increased from $T_C$ = 250–450 C and then decreased to $T_C$ = 550 C (Fig. 3). For Cu – ZnO nanoparticles, the particle size seems to be dependent on the annealing temperature. It varied from 15.86 to 24.25 nm when the annealing temperature was in the range of $250^0$ – $550^0$ C. The increase in the particle size with the annealing temperature (Fig. 3), may be due to the agglomeration of the particles. The increase in the particle size with the annealing temperature as shown in Fig. 3.

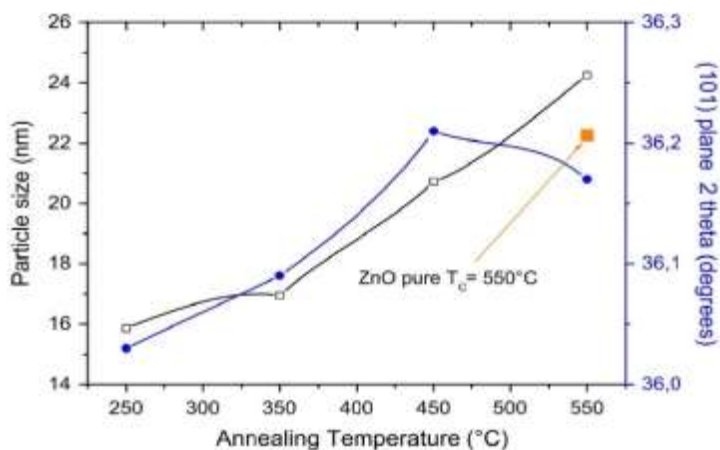

Fig. 3: Effect of annealing temperature on the crystallite size and the (101) plane two thetas.

*3. Fourier transform infrared studies (FTIR)*

The chemical bonding and formation of wurtzite structure in ZnO (pure) and Cu-doped ZnO were confirmed by FTIR measurements at room temperature. The spectra are shown in Fig. 4.The broad absorption band at 3439.39, 1077.74, 3446.53 and 1075.74 cm$^{-1}$ can be attributed to the normal polymeric O-H stretching vibration of $H_2O$, respectively in ZnO and Cu - ZnO lattices [22].Other sharp peaks observed at 1621.45 and 1615.60 cm-1can be attributed to H–O–H bending vibration, which in turn can be assigned to the small amount of $H_2O$ in the ZnO and Cu – ZnO nanocrystals [23].The vibration band at446.31 cm$^{-1}$assigned to the stretching mode of pure ZnO, shifted to a lower frequency at 438.48 cm$^{-1}$ for Cu - ZnO. Hence, it is supposed that Cu-ion can be successfully substituted into the crystal lattice of ZnO [24].

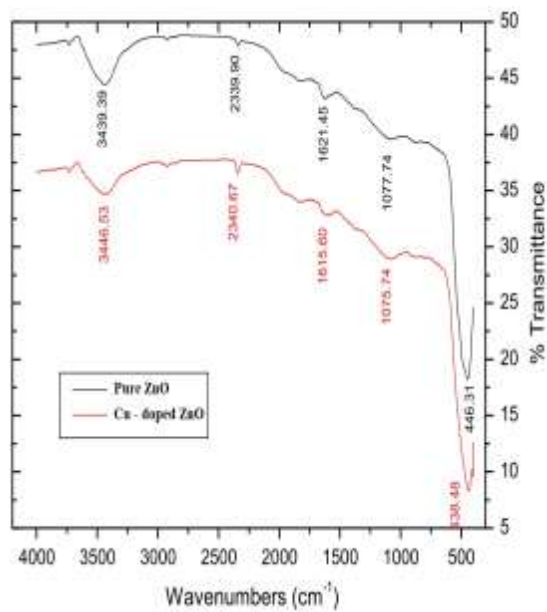

Fig.4: FTIR spectra of pure ZnO and Cu-doped ZnO annealed at $550^0$C

*4. Surface area analysis (BET)*

Figure 5 shows the nitrogen adsorption-desorption isotherms and Barett Joyner–Halenda (BJH) pore size distribution for pure ZnO and Cu-doped annealed at $550^0$C. All the isotherms obtained are Type IV and correspond to a capillary condensation, according to the classification of the International Union for Pure and Applied Chemistry (IUPAC). The hysteresis is Type $H_3$ and is characteristic of the followed by a sharp rise from 0.6 or 0.8 and above due to substantial interparticle porosity [21]. The highest and least volume of absorbed $N_2$ can be attributed to the catalysts annealed at TC equal to $250^0$ and $550^0$ C, respectively. All the desorption branches are different from those of adsorption, indicating differences in their pore's texture. Among all the catalysts, Cu -ZnO annealed at $T_C = 250^0$ C showed the highest volume of adsorption and widest desorption branch as shown in Fig. 5.

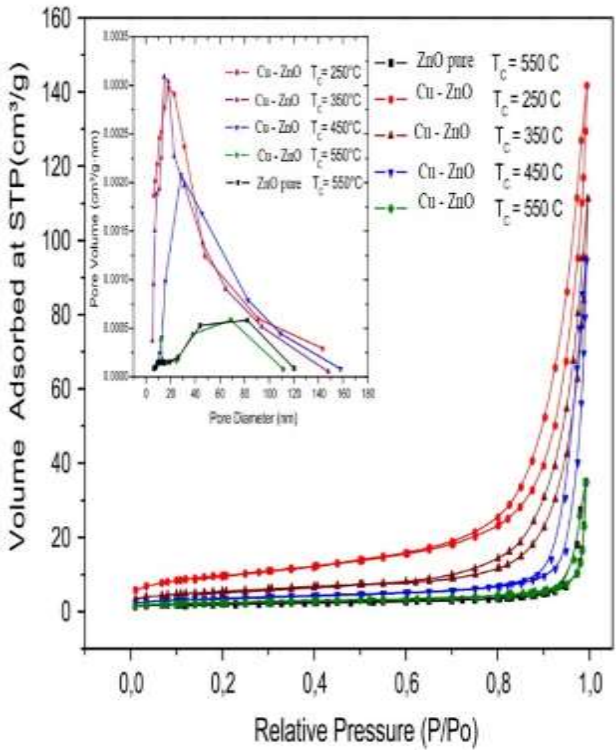

Fig.5: $N_2$ adsorption-desorption isotherms and pore size distribution (inset) of Pure ZnO and Cu-doped ZnO annealed at different temperature

### 5. Characterization of pure ZnO and Cu – ZnO by Photocatalytic degradation of Malachite Green

The photocatalytic performances of pure ZnO and Cu-doped ZnO samples were investigated under UV light through the degradation of Malachite Green (MG), as a model organic pollutant. The photodegradation activity of ZnO annealed at 550 $^0$C and Cu-doped ZnO annealed at TC of 550 $^0$C photocatalysts was shown in Fig. 6. The results show that there is no degradation in the absence of the photocatalyst. On the other hand, the photocatalytic activity of pure ZnO was found to be very low: only 20 % of MG was degraded after 3h of irradiation. However, after doping ZnO with copper significantly increased its photocatalytic efficiency towards the degradation of MG. However, the efficiency of ZnO was significantly increased after doping with copper. As observed in Fig. 6, the kinetics were well fitted to pseudo first order. It is important to note that the photocatalytic efficiency towards the degradation of MG with Cu – ZnO increased with increase in annealing temperature and formation of a CuO phase on the surface. This clearly demonstrates that photocatalytic activity depends on the annealing temperature rather than particles size and surface area. Similarly, it was shown that the photocatalytic activity of a catalyst is related to its microstructure, such as crystal plane, crystallinity, surface properties, BET specific surface area [25]. In this study, the increase in photocatalytic activity of Cu - ZnO $T_C$ = 550 C may be attributed to the good crystallinity and oxygen defects as reported for pure ZnO [26]. However, the increasing crystallinity level with annealing temperature and the introduction of the native defects in the catalyst crystal in the form of neutral (VO), singly charged (VO$^+$) or doubly charged (VO$^{++}$) oxygen vacancies at higher annealing temperatures, may play a major role in the enhancement of its photocatalytic efficiency [27].

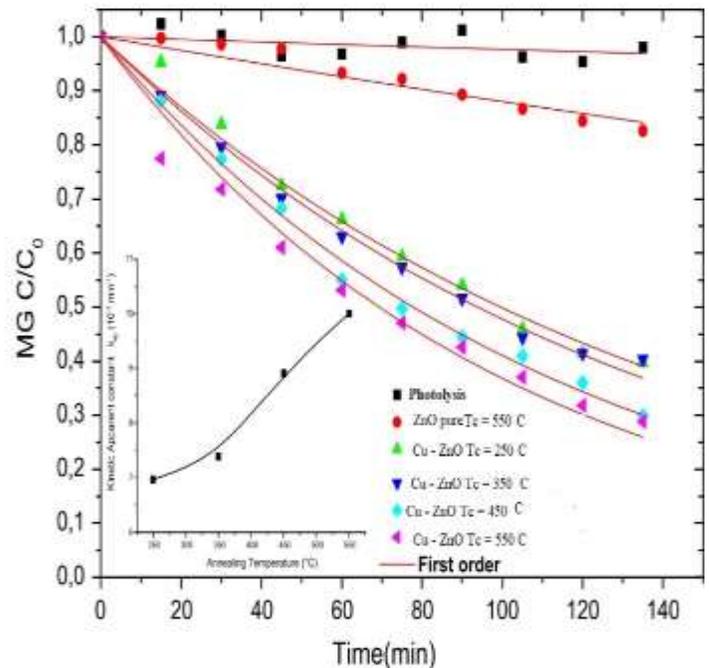

Fig.6: Photodegradation of MG under UV irradiation in the presence and absence of mesoporous catalysts.

## 6. Photocatalytic degradation of MG mechanism

When mixtures of MG aqueous solution and a suspension of photocatalyst were irradiated with VU light, the green solution of MG markedly changed, as a result of the decomposition of the free radicals formed in the solution. The photocatalytic activity mechanism of Cu-ZnO can be understood as follows: The Cu doped ZnO resulted in the creation of intermediate energy levels, which cause a delay in the recombination of charge carriers, thereby enhancing the photocatalytic activity. Moreover, it can be noticed that the delay in the recombination of charge carriers increases whenever the crystallinity of Cu –ZnO photocatalyst is improved by the annealing temperature rise. In the mechanism of photodegradation of MG in the presence of Cu –ZnO photocatalyst. The excitation of photocatalyst by UV light results to the formation of electron $e_{CB}^-$ in both Cu/ZnO and CuO conduction bands. The electrons $e_{CB,CuO}^-$ are transferred to the Cu/ZnO conduction band and then convert dissolved oxygen to super oxide radical. $O_2^-$. While the holes $h_{VB,ZnO}^+$ and $h_{VB,CuO}^+$ from the CuO valence band level react with water to form the strongest oxidizing agent, hydroxide free radical .OH. Both active species can decomposed MG completely.

## 7. Optical Properties

The optical transmission spectra of all samples in the visible region is a very important factor in many applications. So, measured at room temperature are illustrated in Fig. 7. All transmission spectra indicate sharp absorption edge. Moreover, as Cu composition increases, the transmission spectra exhibits the obvious blue shift of absorption edge of an optical band gap of the film and highly transparent within a visible region with increasing Cu composition. The corresponding band gap values of all CZO films can be calculated by the equation:

$$\alpha h v = A \sqrt{h v - E_g} \qquad (8)$$

Where $E_g$ is the optical band gap energy, $A$ is a constant having values between:

$1 \times 10^5 cm^{-1} ev^{-1}$ to $1 \times 10^6 cm^{-1} ev^{-1}$, $hv$ is a photon energy, $\alpha$ is an absorption coefficient.

when $Cu$ composition increases, the transmission spectrum exhibits an obvious blue shift in absorption edge of an optical band gap of the film and better transparency within a visible region with more than 90-100% in its transparency. The band gap of the CZO can be varied from 3.17 to 3.36 eV by varying Cu content meanwhile band gap of the CZO can be increased to 3.36 eV by varying Cu content up to 20 % as shown in Fig.8.

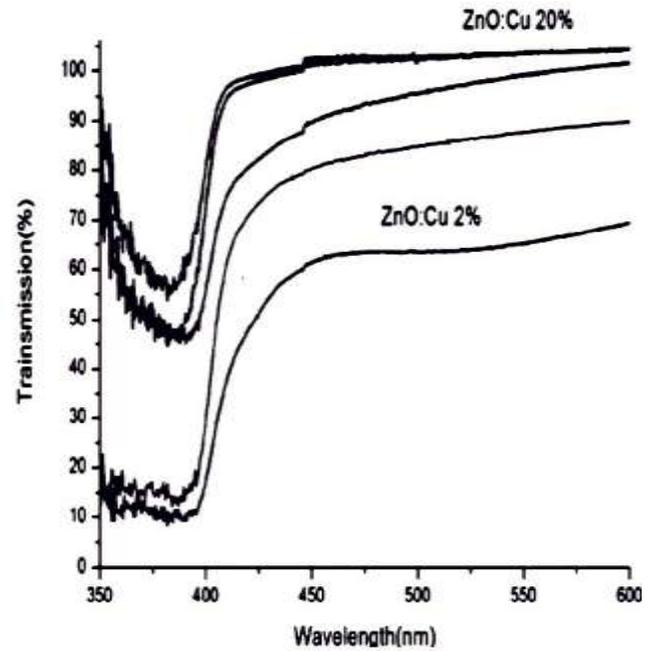

Fig.7: Optical transmittance spectra of Cu-doped ZnO films

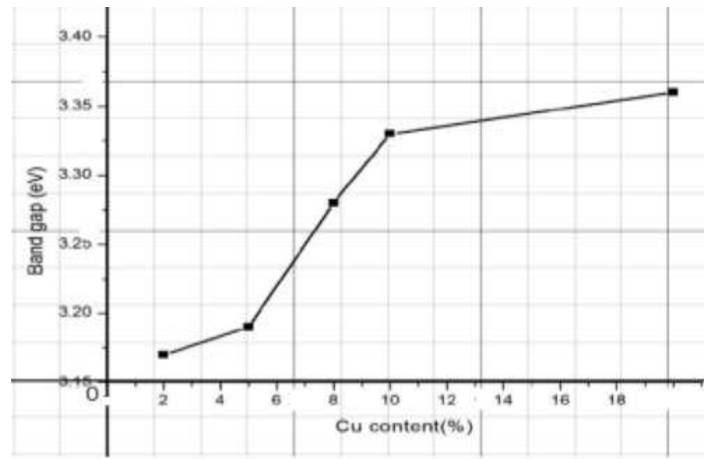

Fig.8: The calculated optical band gap of Cu-doped ZnO films.

## 8. Photoluminescence

The technique of photoluminescence excitation has become a standard one for obtaining information on the nanostructures. The photoluminescence excitation technique involves scanning the frequency of the excitation signal and recording the emission within a very narrow spectral range. In this study, the photoluminescence (PL) spectra of ZnO nanoparticles were determined to explore the effect of Cu-doping and annealing temperature on its optical properties. Figure 9 shows the emission spectrum of pure ZnO and Cu - ZnO catalysts at room temperature. As the annealing temperature increased at $550^0$c, all UV and visible luminescence also increased. The increase in UV emission may be ascribed to the improvement in observed crystalline quality due to annealing. The enhancement in blue emissions intensities is likely due to the strong exchange interactions between Zn and the second phase formed by Cu, after annealing of the sample. The first one, which originates from the recombination of free exciton [28], is clearly observed for pure ZnO at 379 nm. Thus, Cu -ZnO was annealed at the highest temperature TC = 550 C at 385 nm, and this may be attributed to exciton-related near-band edge emission (NBE). The second peak was observed according to the annealing temperature TC at 417, 421, 428.5 and 430 nm, which corresponds to the blue emission [29]. The observation of visible emissions may be related to intrinsic defects in ZnO, and their enhancement in the presence of Cu ions induced a poorer crystallinity and greater level of structural defects, which can be attributed to the more intrinsic defects introduced by Cu ion incorporation into the ZnO lattice. This result is in line with the above XRD findings. Xu et al. [30], related similar results to Raman's observations regarding the defects generated gradually with the Cu - doping ratio.

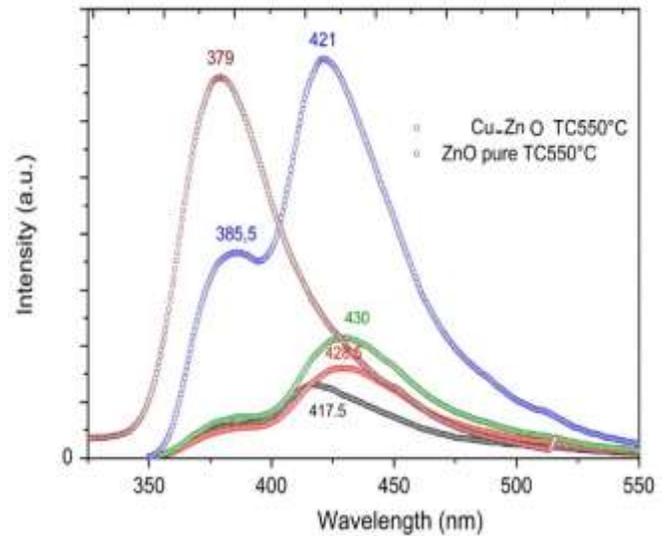

Fig.9: spectra of the pure ZnO and Cu-ZnO annealed at 550 $^0$c.

## 9. Electrical properties of irradiated film

The films samples were treated by $N_2$ ions for various handling times (20, 40, 80 minutes), and the fluence of $N_2$ ions was in the range from $3 \times 10^{17}$ ion/cm$^2$ up to $1.9 \times 10^{18}$ ion/cm$^2$, while the fluence of electrons was in the range from $5 \times 10^{17}$ electron/cm$^2$ to $2.3 \times 10^{18}$ electron/cm$^2$ (Table 1). Electron beams are a shape of ionizing radiation, that means the accelerated electrons have sufficient energy to shatter chemical bonds in organic materials, including polymers. The common effect of the breaking of chemical bonds is the formation of free radicals. Electron beams applications take a worth of processes resulting from the formation of these radicals.

Table 1: Irradiation parameters of film

| Time (minute) | $N_2$ ion beam fluence (ion/cm$^2$) |
|---|---|
| 20 | $3 \times 10^{17}$ |
| 40 | $6 \times 10^{17}$ |
| 80 | $1.9 \times 10^{18}$ |
| Time (minute) | Electron beam fluence (electron/cm$^2$) |
| 20 | $5 \times 10^{17}$ |
| 40 | $9 \times 10^{17}$ |
| 80 | $2.3 \times 10^{18}$ |

Figure 10 (a) shows the variation of log ω versus ln σ for all irradiated and non-irradiated film specimen with electron beam at room temperature for different treatment times (0, 20,40, and 80 minutes). The conductivity increases approximately with increasing frequency, where the change of conductivity with frequency is better as ln σ versus log ω (ω = 2 π f) [31]. Figure 10 (b) shows the same behavior, but for ion beam irradiation on film samples of different doses at room temperature. It is shown that conductivity increases approximately with increasing frequency, as seen for ion beam irradiation (Fig. 1(b)), but with a higher effect than for the electron beam.

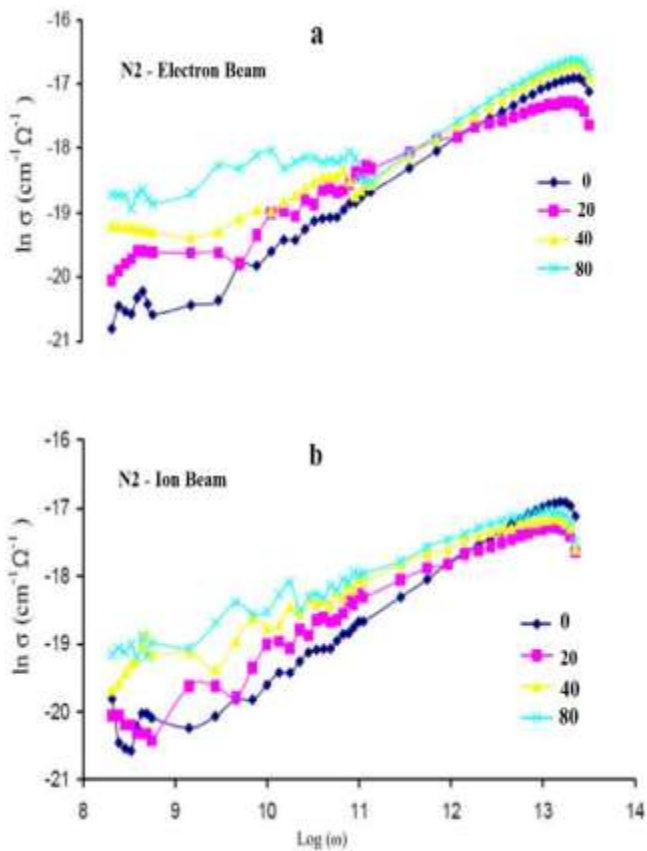

Fig10: Variation in AC lnσ (cm$^{-1}$Ω$^{-1}$) against log ω for all irradiated and non-irradiated samples with electron and ion beams at room temperature

Figure 11(a) shows the variation of the dielectric constant ($\varepsilon'$) of films samples before and after electron and ion beam irradiation with different treatment times (20, 40, and 80 minutes) as a function of the frequency at room temperature. All samples show normal dielectric dispersion behavior, where the dielectric constant, ε , decreases with increasing frequency. Figure 11(b) also shows the variation of the dielectric constant ($\varepsilon'$) of film samples before and after ion beam irradiation with different doses (0, 20, 40, and 80 minutes) as a function of frequency at room temperature. All samples also show normal dielectric dispersion behavior, where the dielectric constant decreases with increasing frequency.

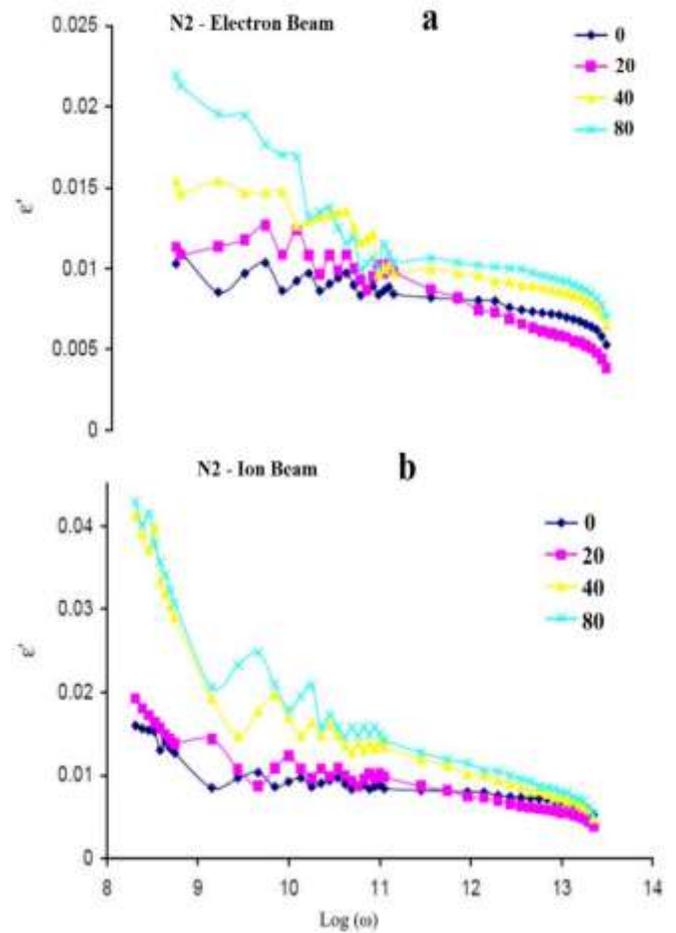

Fig 11: Variation of dielectric loss ε with ln ω of film for all irradiated and non-irradiated samples with electron and ion beams at room temperature.

Figure 12(a) shows the variation of the dielectric constant ($\varepsilon''$) of film samples before and after electron beam irradiation with different treatment times (0, 20, 40, and 80 minutes) as a function of the frequency at room temperature. Figure 12(a) shows the variation of the dielectric constant ($\varepsilon''$) of PVA samples before and after electron beam irradiation with different doses (0, 20, 40, and 80 minutes) as a function of frequency at room temperature. The dielectric constant ($\varepsilon''$) can be evaluated as a function of frequency, where the curves indicate a considerable decrease with a frequency side. A large influence for the ion beam effect on film samples with different doses more than for electron beam (Fig. 12(b)).

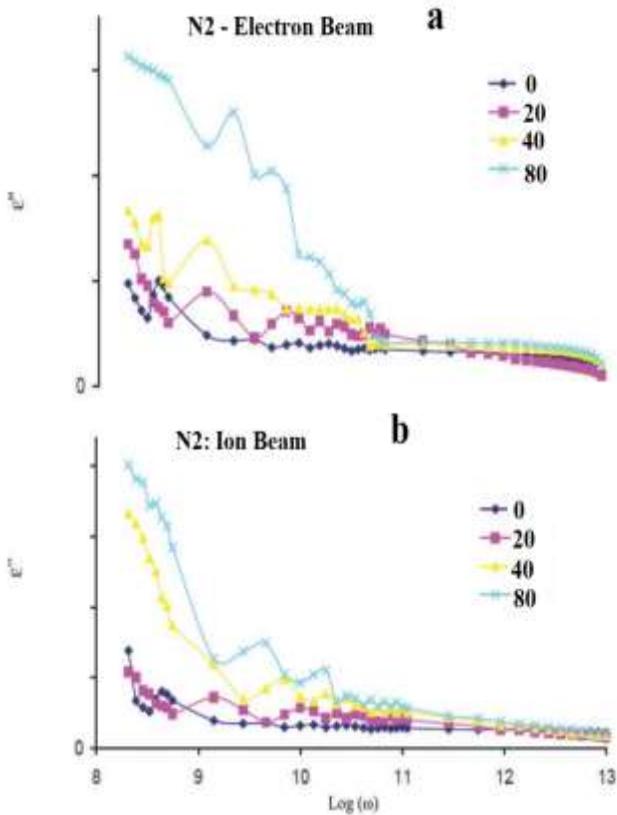

Fig 12: Variation of the imaginary part of the impedance $\varepsilon''$ with frequency for film at room temperature, irradiated and non-irradiated with electron and ion beams.

The dielectric loss tangent (tan δ) as a function of frequency was studied at room temperature (Figs. 13(a), (b)) for ion/electron beam irradiation on film samples with different treatment times (20, 40, and 80 minutes). The dielectric loss tangent decreases with increasing frequency for all irradiated and non-irradiated samples [32,33]. The values of tan $\varepsilon'$ depend upon a number of factors, such as film content and structural homogeneity. From this figure, it can be seen that the dielectric loss tangent has a sudden decrease at low frequency in comparison to the high frequency region (Figs. 13(a), (b)). The influence of ion beam irradiation on films was found to be less than the electron beam effect.

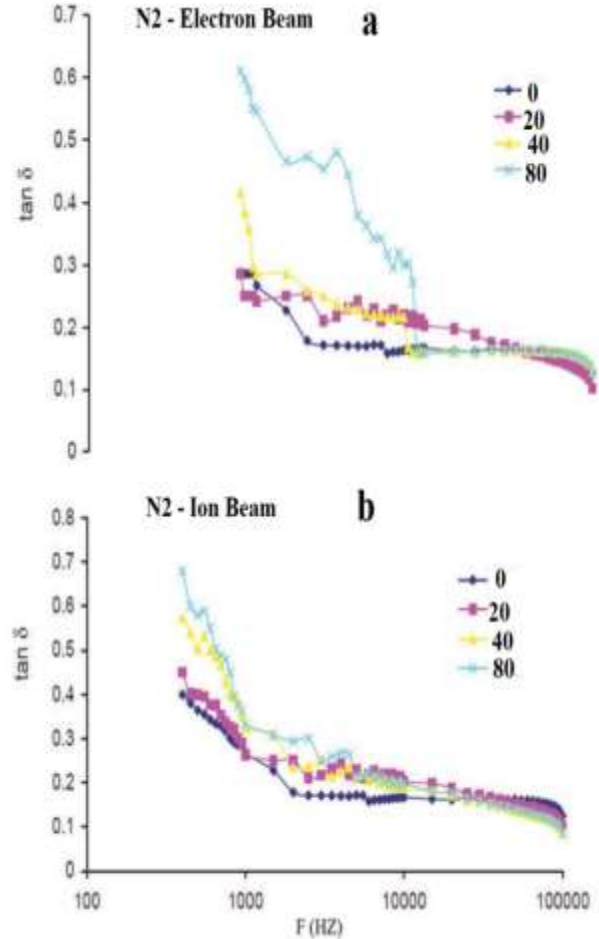

Fig 13: Variation in the dielectric loss tangent tan δ with frequency of film irradiated and non-irradiated with electron and ion beams at room temperature.

## IV. CONCLUSION

One of the more significant findings to emerge from this study is that the structural and optical properties of Cu doped ZnO films prepared by sol-gel method indicates that their properties highly depend on the Cu doping composition of the films. All films are highly transparent and consisted of crystallized ZnO with hexagonal wurtzite structure. With increasing Cu doping content, the crystallinity of the film decreases but band gap of the films inversely increases by varying Cu content. The determined optical parameters are in good agreement with previously reported results. Photoluminescence spectra, showing strong and sharp near band edge emission temperature. This study has shown that its specific surface area lowering with annealing temperature due to the release of the gathering phenomenon. The photocatalytic ability showed that the catalyst activity was influenced by both Cu loading on ZnO and annealing temperature. One of the more significant findings to emerge from this study is that the Cu-doped ZnO catalyst annealed at 550 °C showed the highest photocatalytic activity due to the interlacing of several factors such as the efficient charge separation as proven by the PL spectra, the enhancement of crystallinity and the introduction of the native defects in the catalyst crystal in the form of neutral (VO), singly charged (VO$^+$) or doubly charged (VO$^{++}$) oxygen vacancies. Moreover, the CuO phase on the surface resulted in delayed recombination charge and increased the photocatalytic activity of catalysts at a higher temperature. The influence of an electron/ion beam on film samples with different treatment times (20, 40, and 80 minutes) has been systematically investigated in detail from the viewpoint of the electrical properties. It was found that the dielectric permittivity, and tan δ are found to increase with the irradiation time. There was an increased effect for ion beam compared to electron beam irradiation. The decrease in surface conductivity on (ion /electron) irradiation in film samples is attributed to a decrease in the mobility of macromolecular charged species due to an increase in the degree of crystallinity. Also, electric conductivity increases approximately linearly with increasing frequency, as seen for both electron and ion beam irradiation with different doses, but with a higher effect for ions than for electrons.